\documentclass[aps,epsfig,preprint]{revtex4-1}
\usepackage[section]{placeins}        
\usepackage{float}
 \usepackage{graphics}
 \usepackage{multirow}
\usepackage{caption}
\usepackage{epsfig}
\usepackage{pdfpages}
\usepackage{subcaption}\usepackage{ragged2e}
\usepackage{amsmath,amsthm, amssymb, latexsym}

\newcommand{\be}{\begin{equation}}
\newcommand{\ee}{\end{equation}}

\newcommand{\bea}{\begin{eqnarray}}
\newcommand{\eea}{\end{eqnarray}}

\begin{document}
\title{Effect  of finite beam width on current separation in beam plasma system: Particle-in-Cell simulations}
 \author{Chandrashekhar Shukla}
 \author{Amita Das}\email{amita@ipr.res.in}

 \affiliation{Institute for Plasma Research, Bhat , Gandhinagar - 382428, India }
  \author{Kartik Patel} 
 \affiliation{Bhabha Atomic Research Centre, Trombay, Mumbai - 400 085, India }
\date{\today}
\begin{abstract} 
The electron  beam propagation in a plasma medium is susceptible to  several instabilities. In the relativistic regime typically the weibel instability 
leading to the current separation dominates. The linear  instability 
analysis is carried out for a system wherein the  transverse  extent of the beam is infinite.  Even in simulations, 
infinite transverse extent of the beam has been chosen. In real situations, however,  beam width will always be finite. keeping this in 
view the role of finite beam width on the evolution of the beam plasma system has been  studied here using Particle - in - Cell simulations. 
It is observed that the current separation between the forward and return shielding current for a beam with finite beam 
occurs at the scale length of the beam width itself. Consequently the magnetic field structures that form have maximum power at the 
scale length of the beam width. This behaviour  is distinct from what happens with a  beam with having an infinite extent 
represented by simulations in a periodic box,  where the initial scale length at which the magnetic field appears is that of the electron skin depth. 

\end{abstract}
 \maketitle 
\section{Introduction}
The beam plasma system is of interest to a wide variety of frontline research topics ranging from laboratory to astrophysical plasmas. 
One of the important issue is related towards the understanding of the generation and evolution of magnetic fields in such systems. 
Laboratory experiments conducted with the help of lasers, have shown the generation of very large magnetic fields of the order of Mega Gauss, as the 
energetic electron beam created at the critical surface of the plasma in the target moves inside the bulk plasma \cite{stamper,ar,mb}. The evolution leads to 
eventual turbulence in magnetic field structure. The spatial profile of the turbulent magnetic field intensity have recently been observed experimentally \cite{smondal}. 
The understanding of the generation of the magnetic field is premised on the fact that the energetic electron beam current induces neutralizing 
 shielding current in the plasma through the background electrons. The combined system of beam and  the reverse shielding current is susceptible to a host 
 of beam plasma instabilities. One of the prominent Weibel instability\cite{weibel} is responsible for the spatial separation of current transverse to the propagation direction. 
 The current separation then causes the magnetic field to develop and its evolution subsequently then is believed to be governed by the Electron 
 Magnetohydrodynamic (EMHD) model. At later times when the slow ions would also be ready to participate in the dynamics,  the evolution can be expected to be 
 governed by the Magnetohydrodynamic (MHD ) model.  
 
The Weibel growth rate evaluated for an infinite extent beam is known to  typically maximize at the  electron skin depth scale\cite{lg,ms,mt}. Thus,  one expects the initial 
magnetic field configuration to have structures at this scale predominantly and the spectral power should also peak at the electron skin depth scale. 
However, in the experiments \cite{smondal} conducted recently, wherein a finite extent laser spot  is used to generate electron beam for propoagation inside the 
target, it was observed that even during very initial time the power observed in the magnetic field maximizes at the typical scale of the 
focal spot and not the electron skin depth. In simulations of the beam plasma system, on the other hand the power has been observed to peak 
at the electron skin depth scale only. There is thus a clear distinction between experiments and the simulation results.  It should, however, be noted 
that these simulations have been performed for a uniform electron beam over a box with periodic boundary conditions, essentially corresponding to a beam 
with infinite extent. In this manuscript we specifically consider  Particle - In - Cell (PIC) 
simulations for  a finite extent beam and show clearly that during initial phase 
the power maximizes at the scale of the beam extent. The results were repeated with a variety of PIC codes and the behaviour of the spectra was identically 
observed in each of the simulations.  In the next section we describe the system used for numerical simulations. Section III shows the 
spectral observations and section IV conclusion and discussions. 

\section{Numerical system}
The schematic of the simulation system for finite beam width have been shown in Fig 1. 
The simulation geometry has been chosen in Cartesian plane (X$\times$Y)
and beam propagates in X direction. The absorbing boundary conditions for fields and reflecting boundary conditions for charge particles have been used in all directions. The area of the simulation box R is
30L$\times$ 30L where L($5\times 10^{-6}$) is skin depth of electron plasma corresponding to density $n_{0}=1.1\times 10^{22} cm^{-3}$.
The 300 $\times$ 300 cells have been chosen in simulation box corresponding to grid size $(\Delta L=0.1L)$ and time step is 0.006 fsec. The total number of electrons and ions per cell chosen for the simulations are 200 each. The ions have the charge and mass of a proton and are kept at rest during the simulation. We initialize the particles with density $n_{0}$ in simulation box and the region from 0 to 6L in X direction has been divided into 3 section. The central part of 0 to 6L region of width  
14L(8L to 22L) with density 0.1$n_{0}$ has been initialize with velocity 0.9c where c is speed of light (see Fig.1).

For infinite beam width the schematic of simulation have bee shown in Fig 2. We chose simulation area of size 40L$\times$ 30L and put vacuum
 from 0 to 5L and from 35L to 40L in beam propagation direction. The periodic boundary conditions have been used for fields and particles both. The particle corresponding to density $0.1n_{0}$ from length of 5L to 11L (X direction) have been initialized with velocity 0.9c.
\section{Observations}
We have shown the snapshots of evolution of magnetic field structure with time and corresponding power spectra for finite and infinite beam width case in fig.3 and 4 respectively.
The magnetic field spectra is taken in transverse direction of beam propagation and in center of magnetic field structure in propagation direction. The magnetic field is normalized by $m_{e}c\omega_{0}/e$ and wave vector is by $\omega_{0}/c$
It is clear from the plots that the electron skin depth is the scale at which the magnetic field  structures appear in the infinite beam case.
On the other hand the structure scale is large when the beam is of finite extent. This is more clearly evident  by the plot of spectral power which 
shows a maximum at electron skin depth scale for the case of infinite beam initially. On the other hand when the beam width is taken to be finite 
the spectra is observed to maximize at the scale of the focal spot. The results in Fig.3 and 4 have been carried out by a PIC code 
  developed in India. As time passes the magnetic field structures is broken in smaller filaments and peaks of power shift towards the higher k in finite beam width case while in infinite beam width case the magnetic field structures become bigger and peaks of power spectra shift towards lower k.
\section{Conclusion and Discussions}
We have studied the the effect of  finite beam size on the generation of magnetic structures in the beam plasma system. We observe that a major role 
is played by finite beam size. Unlike the infinite beam where the magnetic field first appears at the electron skin depth scale, in the case of 
finite extent beam the magnetic field structures have maximum power at the scale of the electron beam. 


\clearpage
\newpage
 \bibliographystyle{unsrt}

\FloatBarrier
\begin{figure}[1]
                \includegraphics[width=0.5\textwidth]{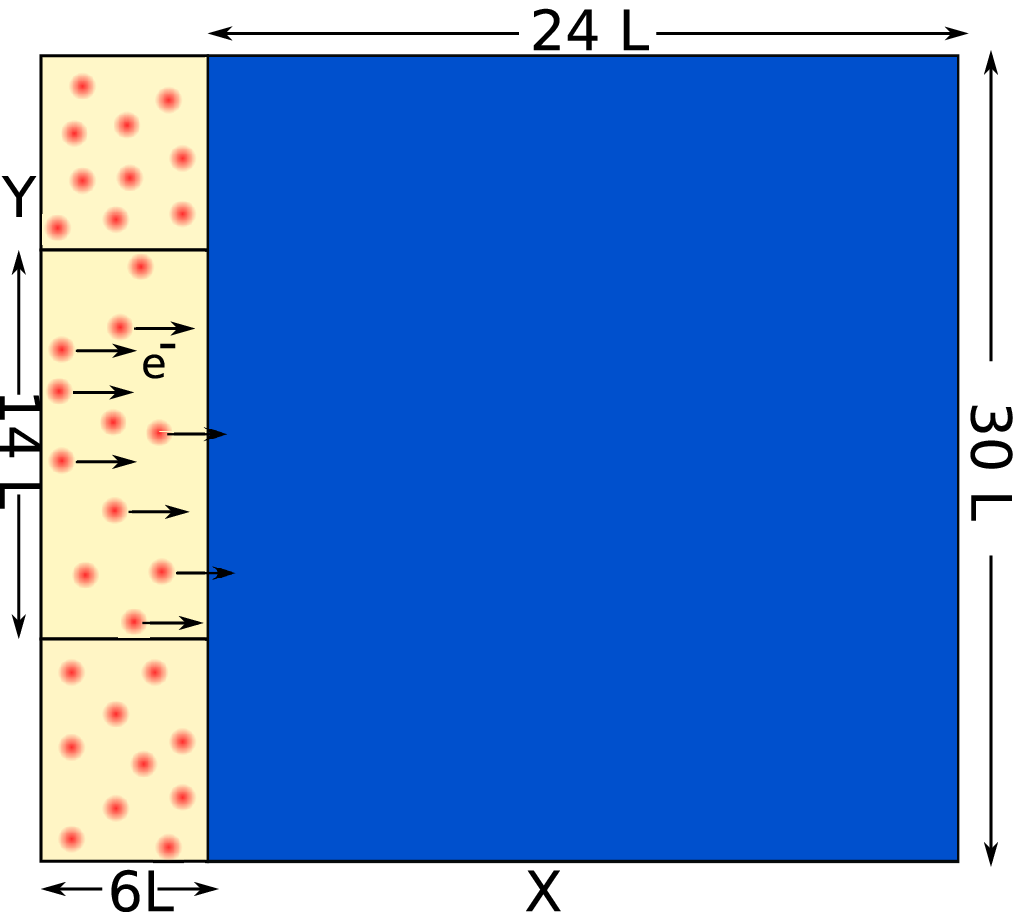}  
              \justifying   \caption{ Sketch of the simulation geometry for finite beam width }  
                 \label{fig:finite}
         \end{figure} 
        \begin{figure}[2]
               \includegraphics[width=0.5\textwidth]{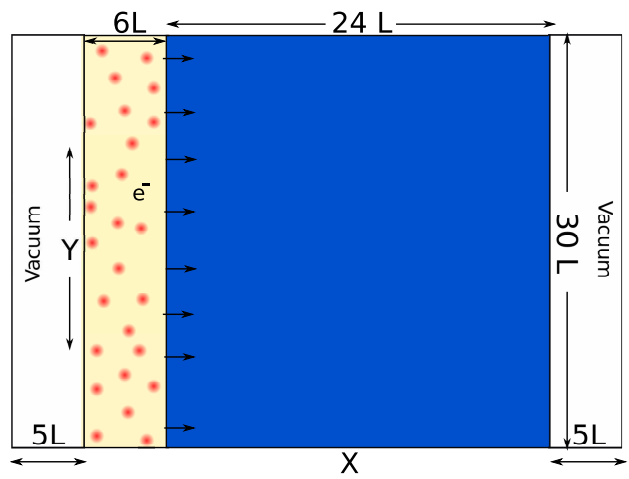} 
              \justifying   \caption{ Sketch of the simulation geometry for infinite beam width with periodic box }  
                 \label{fig:in}
         \end{figure}                  
 \begin{figure}[2]
                \includegraphics[width=\textwidth]{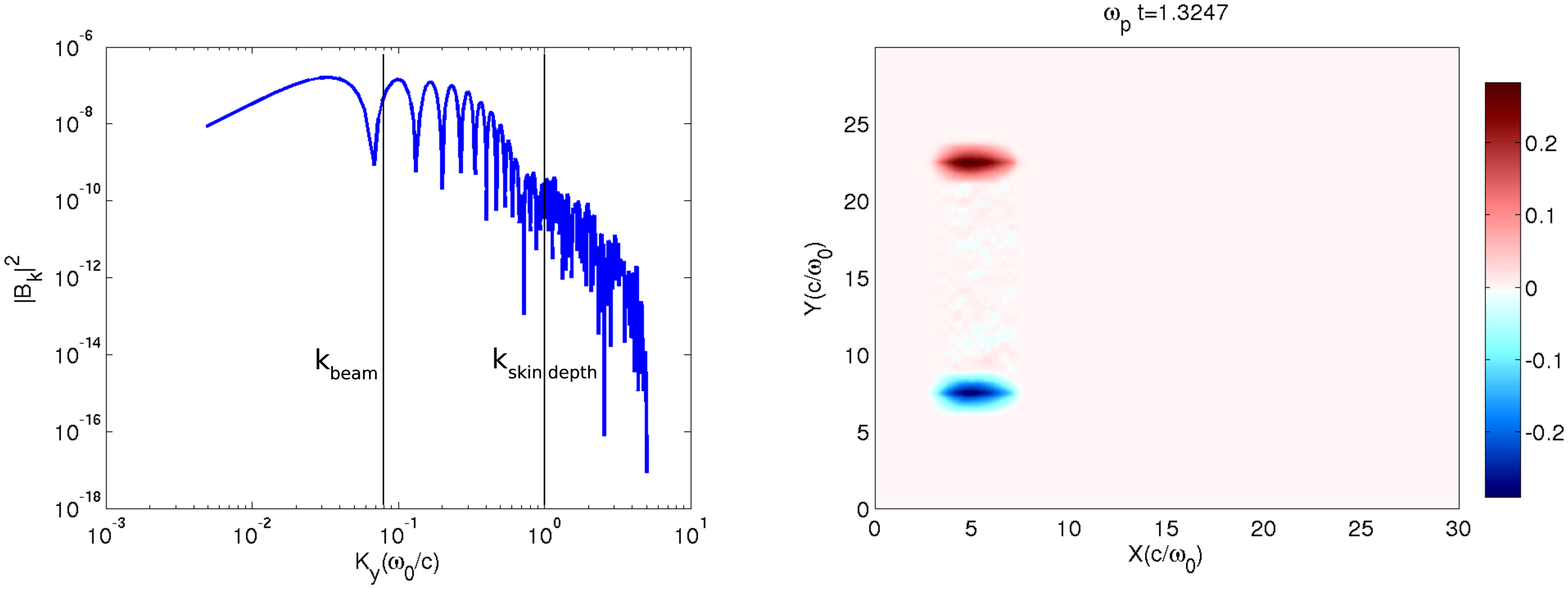}\\
                \includegraphics[width=\textwidth]{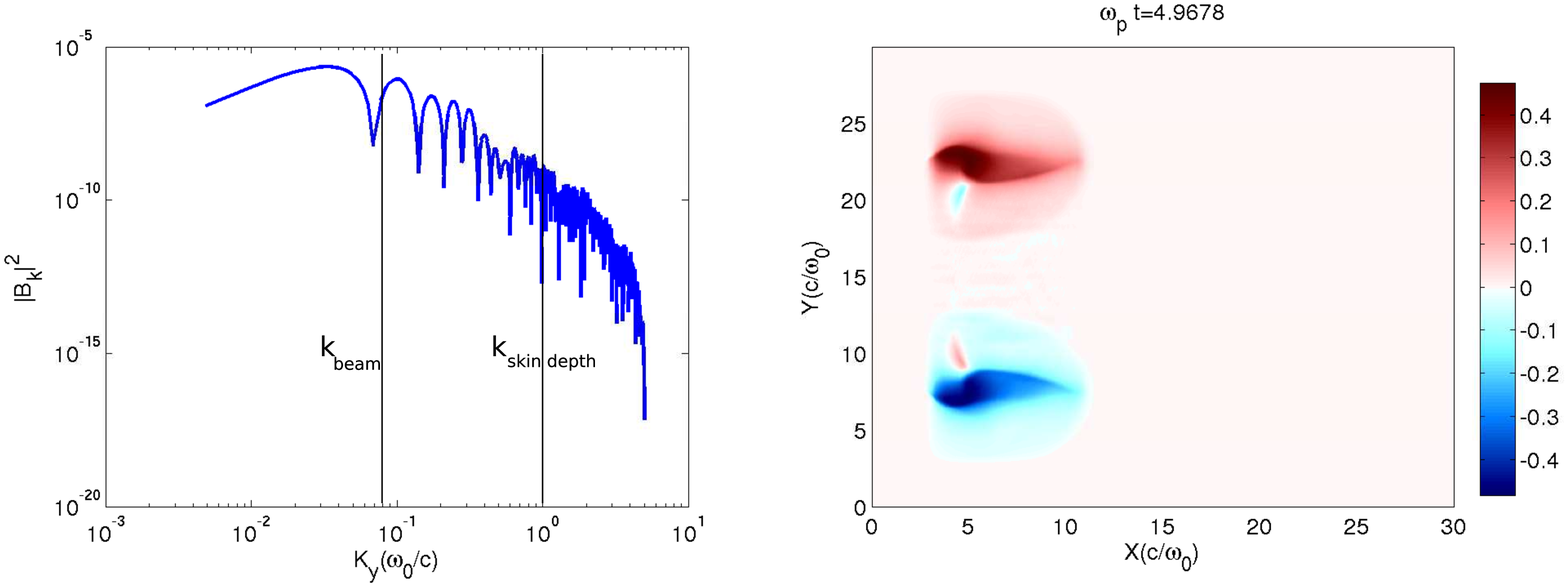} \\
                 \includegraphics[width=\textwidth]{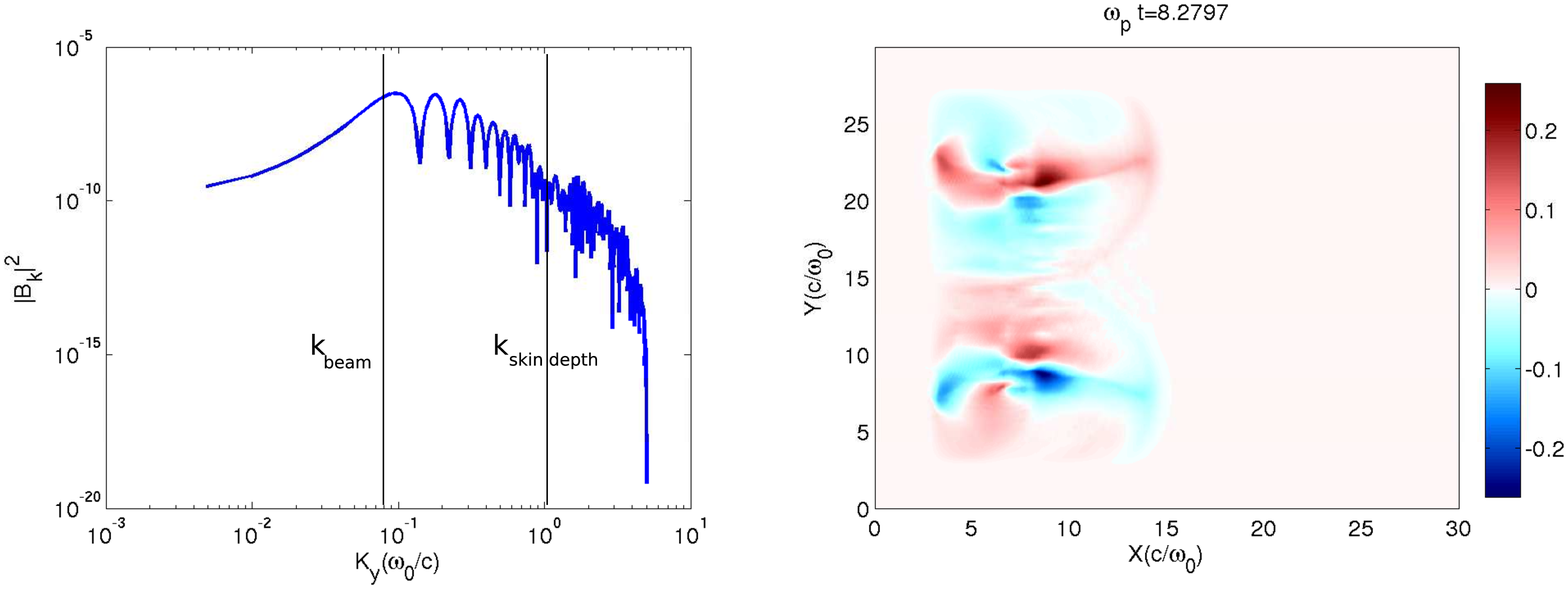} 
              \justifying   \caption{Time evaluation of magnetic field structures and corresponding power spectra for finite beam width case }  
                 \label{fig:finite_field}
         \end{figure}                  
        \begin{figure}[5]
                \includegraphics[width=\textwidth]{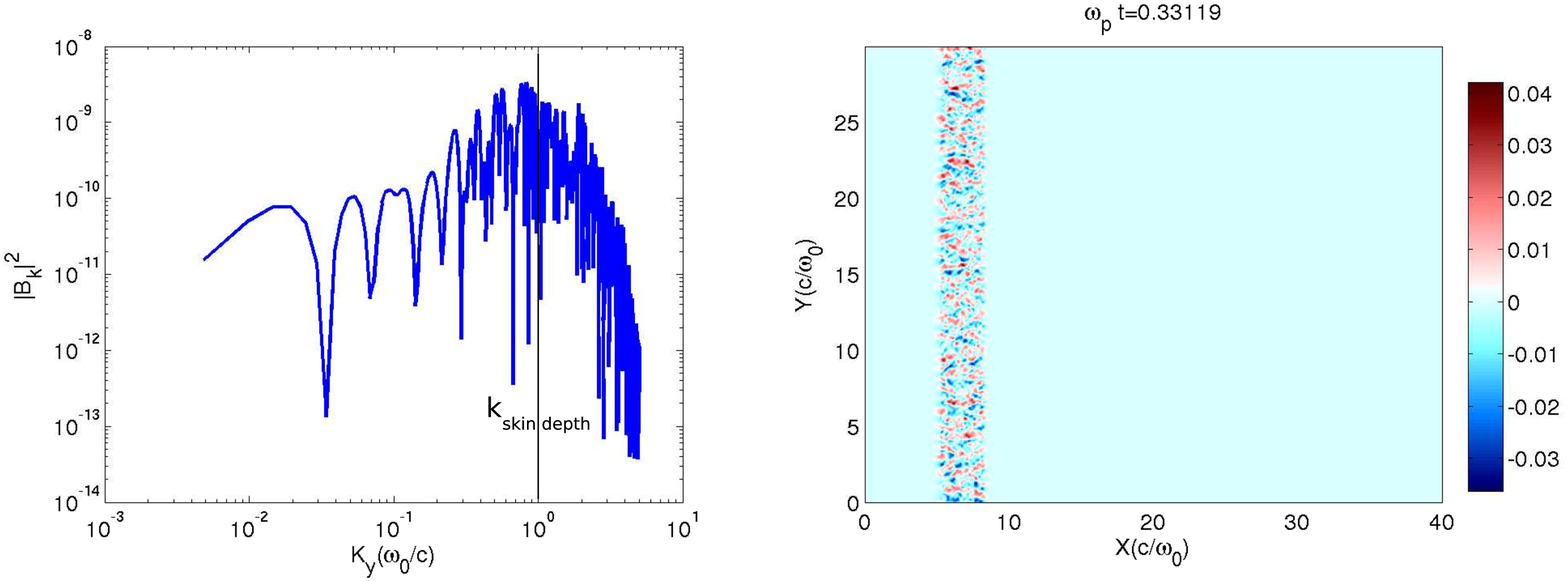}\\
                \includegraphics[width=\textwidth]{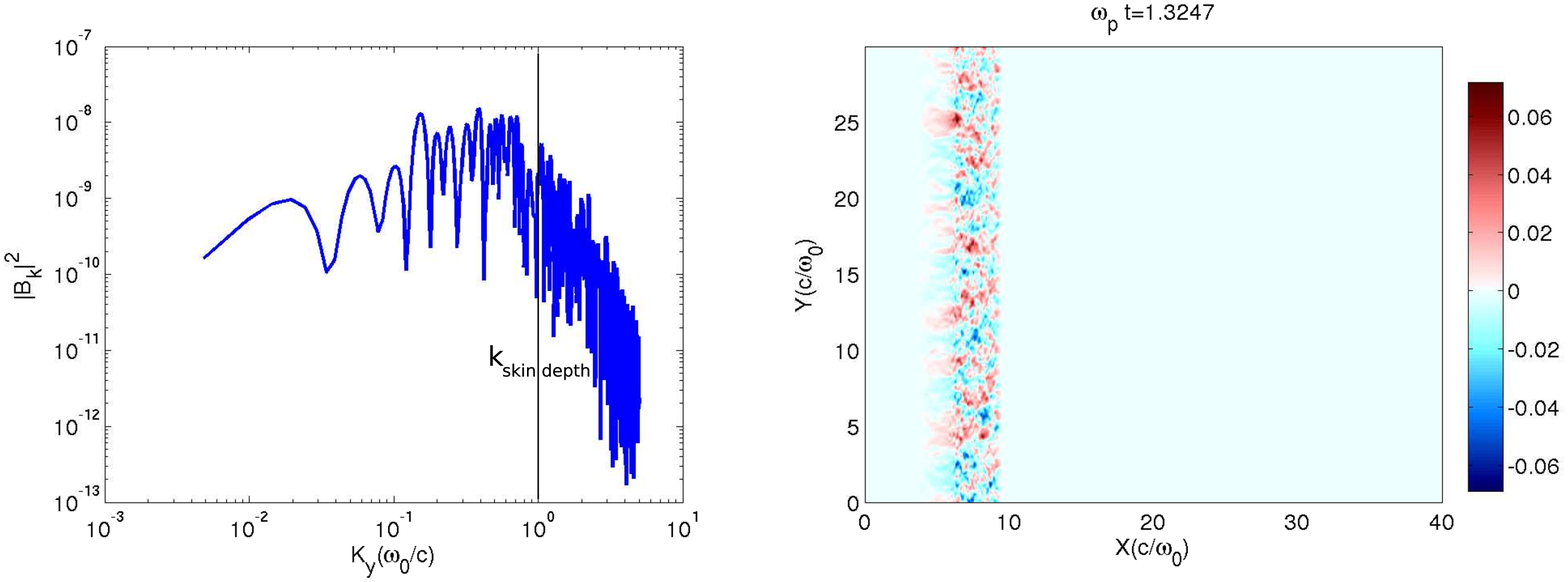} \\
                \includegraphics[width=\textwidth]{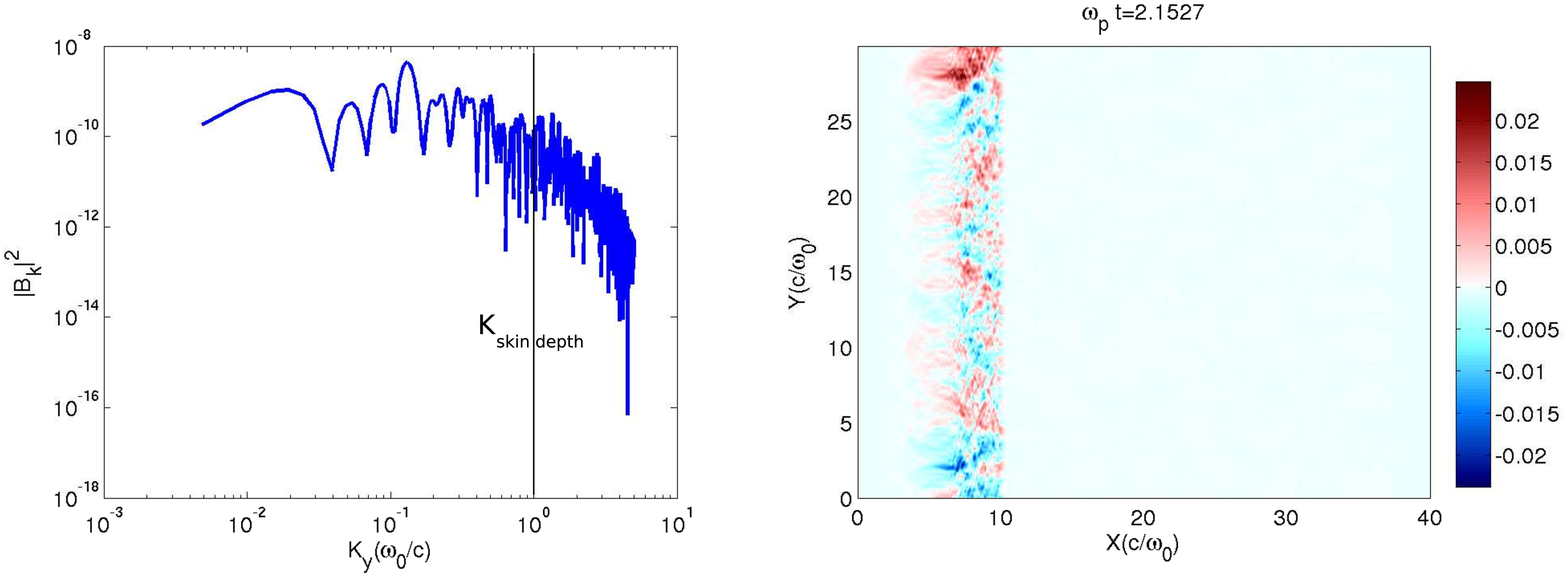} 
               \justifying   \caption{ Time evaluation of magnetic field structures and corresponding power spectra for infinite beam width case}
                 \label{fig:infinite}
         \end{figure} 

\end{document}